\documentstyle[preprint,aps]{revtex}
\begin{document}
\draft
\tightenlines
\title{Tailoring Many-Body Interactions to Solve\\
Hard Combinatorial Problems}
\author{Haiqing Wei}
\address{
Department of Physics, McGill University\\
Montreal, Quebec, Canada H3A 2T8\\
{\rm E-mail: dhw@physics.mcgill.ca}
}
\author{Xin Xue}
\address{
Department of Natural Resource Sciences\\
Macdonald Campus of McGill University\\
Ste-Anne-de-Bellevue, Quebec, Canada H9X 3V9\\
{\rm E-mail: xkhz@musicb.mcgill.ca}
}
\maketitle

\begin{abstract}
A quantum machine consisting of interacting linear clusters of
atoms is proposed for the 3SAT problem. Each cluster with two relevant
states of collective motion can be used to register a Boolean variable.
Given any 3SAT Boolean formula the interactions among the clusters can be
so tailored that the ground state(s) (possibly degenerate) of the whole
system encodes the satisfying truth assignment(s) for it. This relates the
3SAT problem to the dynamics of the properly designed glass system.
\end{abstract}
\pacs{PACS numbers: 89.80.+h, 71.45.-d}

Equipped with powerful algorithms, today's electronic computers solve many
problems amazingly fast. Yet there are problems hard to them in the sense
that the best algorithms essentially take exponential running time. Many of
the hard problems are NP even NP-complete [1]. A problem is NP provided that
it can have its answer checked in polynomial time. If the problem has the
further property that all problems in the NP class can be translated into it
by algorithms taking polynomial time, it is NP-complete [1].

Recently attempts to search for novel machines with greater computational
power are stimulated by the growing evidence that no general efficient
polynomial time solution running in classical computers exists for any
NP-complete problem. DNA computers have been proposed to solve the directed
Hamiltonian path problem [2] and the satisfiability problem [3]. These
proposals took the advantage of the enormous parallelism of solution-phase
chemistry. However, the number of molecules required rises exponentially as
the complexity of the problem increases, which stops the application of the
DNA computation to large scale problems [4,5,6]. Quantum computers may
also solve some problems fast by virtue of the so-called quantum parallelism
[7,8]. But this quantum parallelism seems difficult to harness. So far no
fast quantum algorithm has been found for any NP-complete problem [9].
Furthermore, it is a thorny issue to construct a quantum computer and
maintain the quantum coherence to achieve any useful computation [10,11].
Nevertheless, recent studies reveal the essential role of physics in
computer science. People realize more and more that, what can be computed
and how fast it can be computed is not only a problem of pure computer
science, but a question posed to physics as well.

The satisfiability problem (SAT) is among the first known NP-complete
problems. Consider formulas over a set of Boolean variables
$$V=\{v_1, v_2, ..., v_m\}$$
where each variable can only have values $0$ (false) and $1$ (true). All
variables $v_i$ and their negations $\bar{v}_i$ are called literals over
$V$. A clause is a formula of the form
$$l_1\vee l_2\vee \cdots \vee l_k$$
where each $l_i$ is a literal ($\vee$ is the logical OR operation, $\wedge$
is the logical AND correspondingly). It is said that a clause is satisfied
if and only if at least one of its member literals has the value $1$ so that
the clause gets the value $1$. The famous SAT problem is to ask whether
there is a truth assignment that satisfies a formula of the form
\begin{equation}
F=C_1\wedge C_2\wedge \cdots \wedge C_n
\end{equation}
where each $C_i$ is a clause. The 3SAT problem is just the restricted SAT
problem in which all clauses have exactly three literals.

In this letter, a quantum search machine consisting of linear clusters of
atoms is proposed to encode the 3SAT problem. Each cluster has two relevant
states of collective motion which can be used to register a Boolean variable.
It will be shown that the interactions among the clusters can be so tailored
that the ground state of the whole system encodes an assignment of Boolean
values which solves the satisfaction problem. Later it will be clear that
the restricted structure of 3SAT greatly simplifies the tailoring of
many-body interactions. If in some situation, the system could quickly
relaxes to the ground state, the solution can be read out fast. However,
in the worst cases, the system will be trapped in local energy minima, the
relaxation is exponentially slowed down. Anyway, our model machine provides
a direct connection between an NP-complete problem and a properly designed
glass system. Studying the dynamics of the glass system bears significance 
on learning how to solve the NP problem.

If each Boolean variable is registered by an individual atom or electron, it
won't be easy to tailor the many-body interactions in the desired way.
Specificly for the 3SAT problem, a Boolean variable may enter many clauses, it
is convenient to encode a variable into the collective motion of a chain of
atoms (called a linear cluster of atoms). For example, a one-dimensional
ferromagnetic chain does the favor, the two directions of magnetization of
the chain can be used to store binary information. Lent {\it et al}. have
proposed a binary wire consisting of interacting quantum-dot cells [12,13].
Although this kind of binary wire is by no means the only possible candidate
for implementing the search machine, it serves as a very good illustration. In
this letter, whenever the term binary wire appears, it refers to the wire
proposed by Lent {\it et al}., and more specificly the wire
consisting of {\it rotated cells} [13]. In a binary wire, the two ways of
{\it cell polarizations} are used to encode binary information. A remarkable
advantage of the binary wire system is that it is very easy to achieve logic
fan-out. The quantum machine encoding a simple 3SAT problem
\begin{equation}
F=(\bar{x}_1\vee x_2\vee x_4)\wedge (x_1\vee \bar{x}_3\vee x_4)
\end{equation}
is shown in Fig.1 as an example. There are four Boolean variables which are
registered by four horizontal binary wires labeled by $x_1$, $x_2$, $x_3$ and
$x_4$ respectively. It is assumed that the system is at low enough temperature
so that any free wire stays at the two-fold degenerate ground state, no
higher energy collective mode can be excited. Hence for the four bit
register, the only relevant states are those $2^4$ expanded states called the
working modes which are degenerate. The register is connected by vertical
transmission wires to three-literal clause evaluators (3CEs). Notice these
inverters on some transmission wires (it is easy to implement a logic inverter
[13]). Also notice that both the transmission lines and the inverters conserve
the energy degeneracy among the working modes. The energy degeneracy
conservation will be broken when the 3CEs are connected. Actually the 3CE
is a device inside which the relevant three binary wires are brought close
to interact and the interactions are properly tailored so that the energy
degeneracy among the eight working modes expanded by the three related
literals is lifted according to whether they satisfy the corresponding clause
or not. As long as the clause is satisfied, the energy remain lower, otherwise
the energy is pushed higher. Specificly, for a 3CE connecting to three
literals $(l_1, l_2, l_3)$, the expanded eight states fall into two energy
levels: if at least one of the literals has the value $1$,
the clause $l_1\vee l_2\vee l_3$ is satisfied, the system remains at the
lower energy level; when all the values of the three literals are $0$, the
system will be raised to a higher energy level. A concrete implementation of
the 3CE will be shown later. At this stage let's see how the search machine
works provided that the three-literal clause evaluators behaving as desired
are available. Generally, a 3SAT problem has $m$ Boolean variables and $n$
3-clauses involved (there is an obvious upper bound for $n$, {\it i.e.}
$n<C_{2m}^{3}$). One may construct a quantum search machine with an $m$ bit
register, $n$ 3CEs, a sufficient number of transmission wires, and some
inverters. The working modes associated with the $m$ bit register
consist of the $2^m$ degenerate ground states. Because the register, the
transmission wires, and the inverters conserve the energy degeneracy among
the working modes, while the 3CEs lift the energy degeneracy according to
whether the corresponding clauses are satisfied or not, the whole system of
the search machine will be in its ground state(s) if and only if all 3-clauses
are satisfied. When the system is in the ground state, one measures the
register and gets the solution for the SAT problem if only it exists,
otherwise the measurement result will tell that there is no truth assignment
at all which satisfies all the given clauses.

Now it is time to show how the 3CE can be really implemented. One may
symmetricly arrange the binary wires so that the Hamiltonian for the 3CE
is invariant under any permutation of the three literals. There are eight
states expanded by the three literals $(l_1, l_2, l_3)$ which serve the
labels for the states at the same time. The permutation symmetry divides the
eight states into four classes
$$
\begin{array}{rcl}
class\hspace{1 mm}0:&\hspace{2 mm}&(0 0 0)\\
class\hspace{1 mm}1:&\hspace{2 mm}&(0 0 1), (0 1 0), (1 0 0)\\
class\hspace{1 mm}2:&\hspace{2 mm}&(1 1 0), (1 0 1), (0 1 1)\\
class\hspace{1 mm}3:&\hspace{2 mm}&(1 1 1)
\end{array}
$$
Without the tailored interactions, all these states are degenerate. The
purpose of interaction tailoring is to lift the degeneracy into two energy
levels, keep the seven states in class 1, 2, and 3 at the lower level while
raise the state in class 0 to a higher energy level. The Hamiltonian of the
3CE for three given literals $(l_1, l_2, l_3)$ includes one-body terms
$H_{i}^{(1)}$, two-body terms $H_{ij}^{(2)}$ and a three-body term
$H_{123}^{(3)}$,
\begin{equation}
{\cal H}_{3CE}=\sum _{i}H_{i}^{(1)} +
\sum _{i\neq j}H_{ij}^{(2)} +
H_{123}^{(3)}
\end{equation}
where the one-, two-, and three-body terms are given by
\begin{eqnarray}
H_{i}^{(1)}   &=& 2A\left(\frac{1}{2}-l_i\right) \nonumber \\
H_{ij}^{(2)}  &=& 2B\left(l_i\oplus l_j-\frac{1}{2}\right) \\
H_{123}^{(3)} &=& \sqrt{D^2+\left(E+F\sum _{i=1}^{3}l_i\right)^2}
\nonumber 
\end{eqnarray}
where $i,j=1,2,3$, $i\neq j$, $l_i$ and $l_j$ are Boolean variables that can
have the value $0$ or $1$, $\oplus$ is the exclusive-OR operation, and $A$,
$B$, $D$, $E$, and $F$ are parameters characterizing the many-body
interactions. To make a one-body term $H_{i}^{(1)}$ is straightforward:
applying a bias electric field, its side effect splits the two degenerate
ground states of a binary wire. Notice that $A$ is easily adjustable. It is
also easy to have a two-body
interaction, simply draw the two binary wires $l_i$ and $l_j$ close, the
Coulomb interaction between their charge distributions gives the desired
$H_{ij}^{(2)}$. The only tricky implementation is for $H_{123}^{(3)}$. As
shown in Fig.2, one may place in the same plane, say $(x,y)$ plane, the three
binary wires symmetrically with their ends sitting at a same circle (the
readers may notice at once the fact that there are inevitably two-body
interactions when the three wires are brought close). At a point $P$ out of
$(x,y)$ plane just above the center of the circle, the $z$-component of the
electric field due to the charge distribution of each binary wire $l_i$ is
\begin{equation}
E_i=Fl_i+{\rm constant},\hspace{2 mm}i=1,2,3
\end{equation}
The resultant electric field $E_{tot}$ along the $z$ direction at point
$P$ due to the three wires and an on purpose applied external field with
variable intensity, will be
\begin{equation}
E_{tot}=E+F\sum _{i=1}^{3}l_i
\end{equation}
where $E$ is variable. There is a responser placed at point $P$ which yields
an energy varying according to $E_{tot}$ by the virtue of the Stark effect
[15]. The responser should be sensitive only to the $z$-component of the
electric field. A quantum dot made from a symmetric double well along the $z$
direction can serve as a good responser as desired. In $x,y$ directions the
quantum confinement effect of the dot is very strong leading to large energy
spacing, the Stark effect is negligible. While in the $z$ direction, the
symmetric double well leads to a very good two-level system with relatively
small energy spacing [16] which is sensitive to perturbations. The
Schr\"odinger equation of a two-level system perturbed by an external field
can be exactly solved [17]. The exact eigen-energies are given by
\begin{equation}
U=\pm\sqrt{D^2+\left(pE_{tot}\right)^2}
\end{equation}
where $\pm D$ are the energy levels without perturbation, $p$ is the
transition electric dipole. The combination of Eqs.6 and 7 leads to
the desired three-body energy term $H_{123}^{(3)}$, where for simplicity
the constant dipole is set to $p=1$. Now, according to Eq.3, the total
energies of a 3CE at states of $class$ $0$, $1$, $2$, $3$ are respectively
$U_0$, $U_1$, $U_2$, $U_3$, 
\begin{eqnarray}
U_0 &=&  3A+3B-\sqrt{D^2+\left(E+3F\right)^2} \nonumber \\
U_1 &=&   A- B-\sqrt{D^2+\left(E+ F\right)^2} \nonumber \\
U_2 &=&  -A- B-\sqrt{D^2+\left(E- F\right)^2}           \\
U_3 &=& -3A+3B-\sqrt{D^2+\left(E-3F\right)^2} \nonumber
\end{eqnarray}
In practice, the parameters $B$, $D$, and $F$ are usually fixed, not easy
to adjust, while $A$ and $E$ are ready to vary. It is convenient to measure
all the parameters in unit of $D$. For any given parameters $B/D$ and $F/D$,
The interaction tailoring actually is to adjust the two parameters $A/D$
and $E/D$ so that the following condition holds 
\begin{equation}
U_1=U_2=U_3<U_0
\end{equation}
The equations can be numerically solved. In Fig.3 the points lying in the
black area are of the parameters $B/D$ and $F/D$ for which suitable values of
$A/D$ and $E/D$ (called good 3CE solution) can be found satisfying the
condition (9), under which the seven states in classes $1$, $2$, and $3$ are
degenerate, while the energy of the only state in the class $0$ is lifted
sufficiently higher. Actually the black area in Fig.3 is constricted to the
points for which the stronger condition is satisfied
\begin{equation}
U_1=U_2=U_3\le U_0-0.2D
\end{equation}
The wide black area indicates the fact that for a wide range of $B/D$ and
$F/D$, which are fixed upon the completion of the device fabrication, good
3CEs can be achieved by varying $A/D$ and $E/D$ which are directly related to
adjustable external bias fields. This fact is particularly advantageous to
device fabrication.

The 3CE proposed here is of course not unique. It is only for illustrative
purpose. One may implement 3CEs by other means. The key point is to keep the
states in the classes $1$, $2$, and $3$ degenerate while lift the state in
the class $0$ to a higher energy.

In conclusion, it has been showed that a quantum search machine consisting of
lines of interacting quantum dots can encode the 3SAT problem efficiently in
the sense that the solution is kept in the ground state of the system and the
machine contains a number of 3CEs less than $C_{2n}^{3}$ where $n$ is the
number of Boolean variables. The purpose of the present letter is to show the
possibility of tailoring the many-body interactions in a ``static quantum
network'' to solve some hard computational problems. The example 3SAT problem
is a hard NP problem to classical computers in the sense that with the best
algorithm a classical computer needs exponential running time to find the
solution. The problem becomes intractable when its input size gets larger [1].
To our machine there is the open question of how to make it relax quickly to
the ground state so that it gives us the solution in a short time.

\vspace{2 cm}

\begin{center}
{\large FIGURE CAPTIONS}
\end{center}

\noindent
Fig.1  The search machine solving the simple 3SAT problem
$(\bar{x}_1\vee x_2\vee x_4)\wedge (x_1\vee \bar{x}_3\vee x_4)$.

\noindent
Fig.2  Three binary wires are brought close to yield the
three-body interaction.

\noindent
Fig.3  The area of parameters $B/D$ and $F/D$ where
good 3CE solutions exist. 

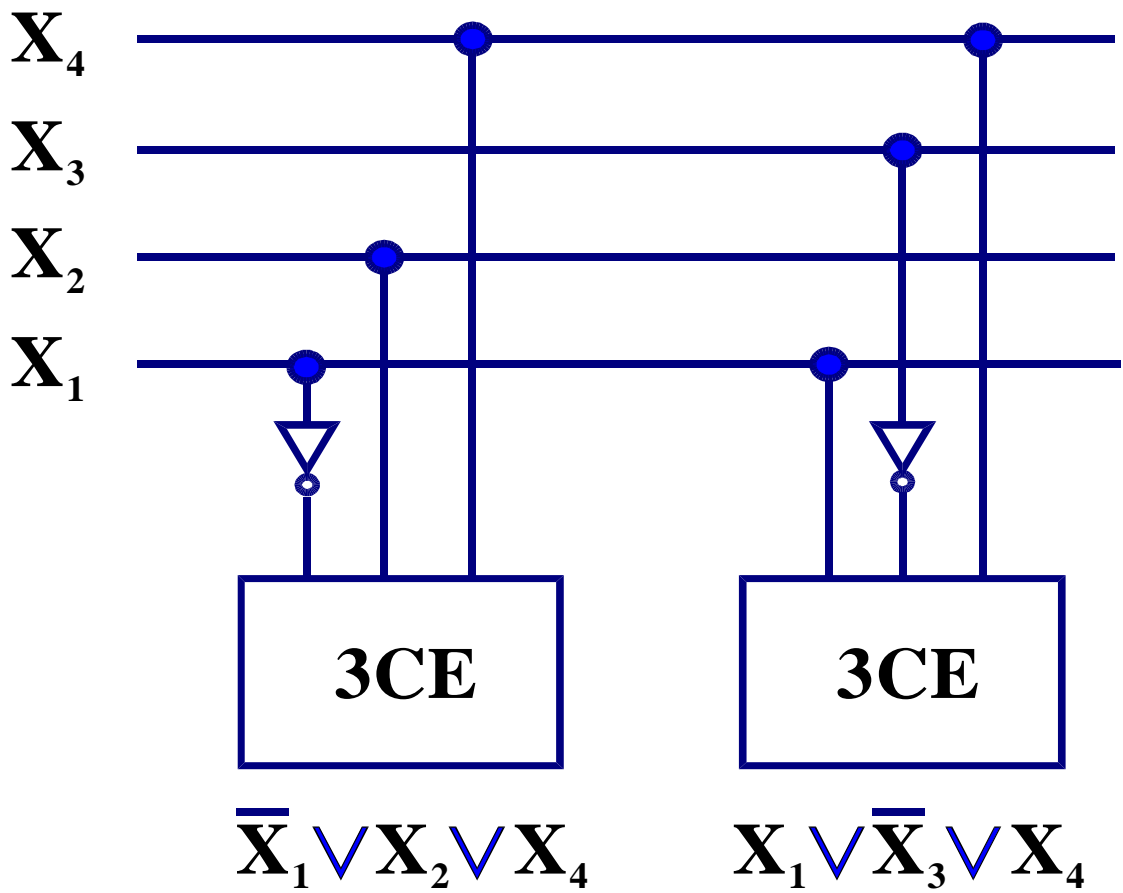
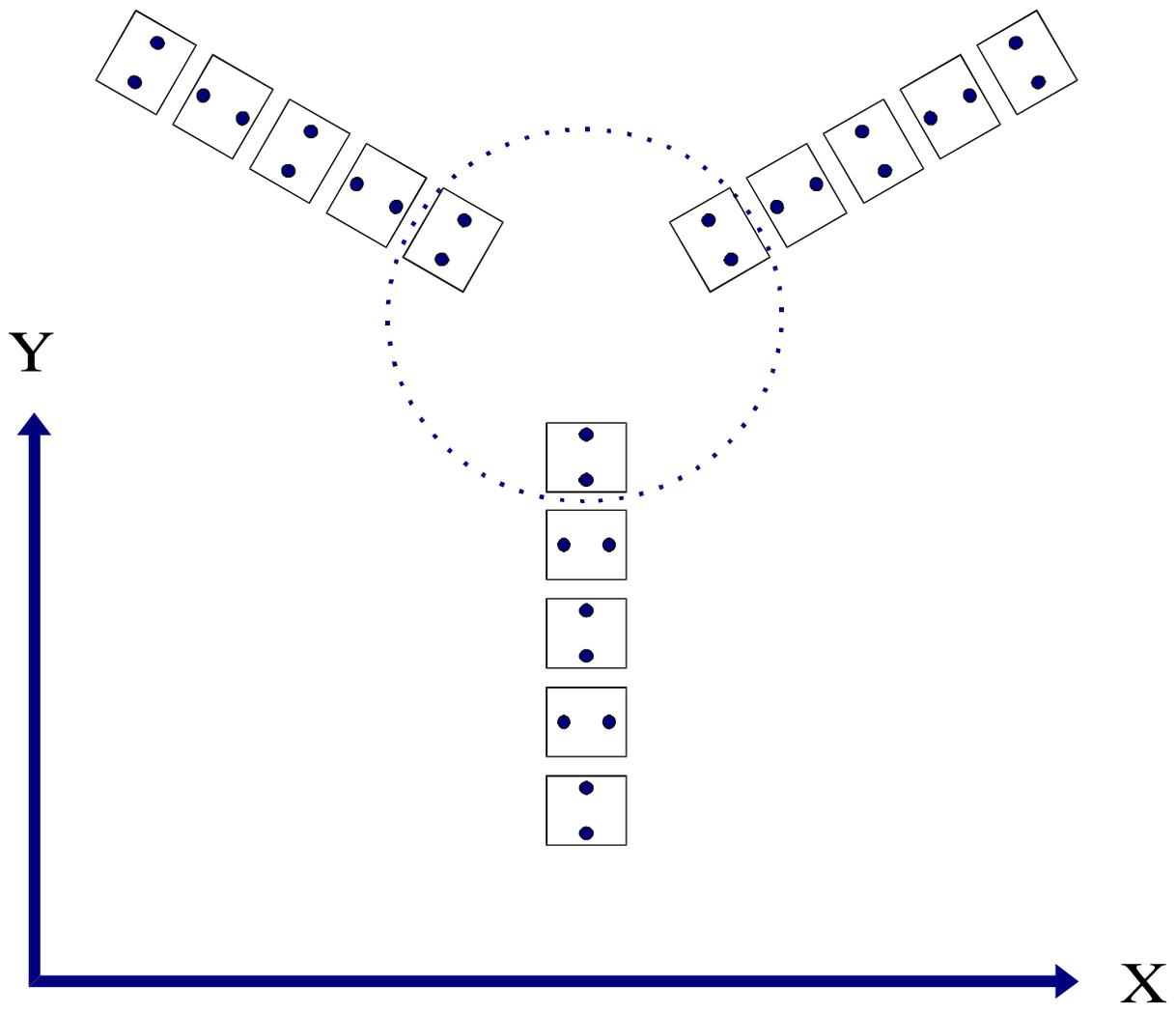
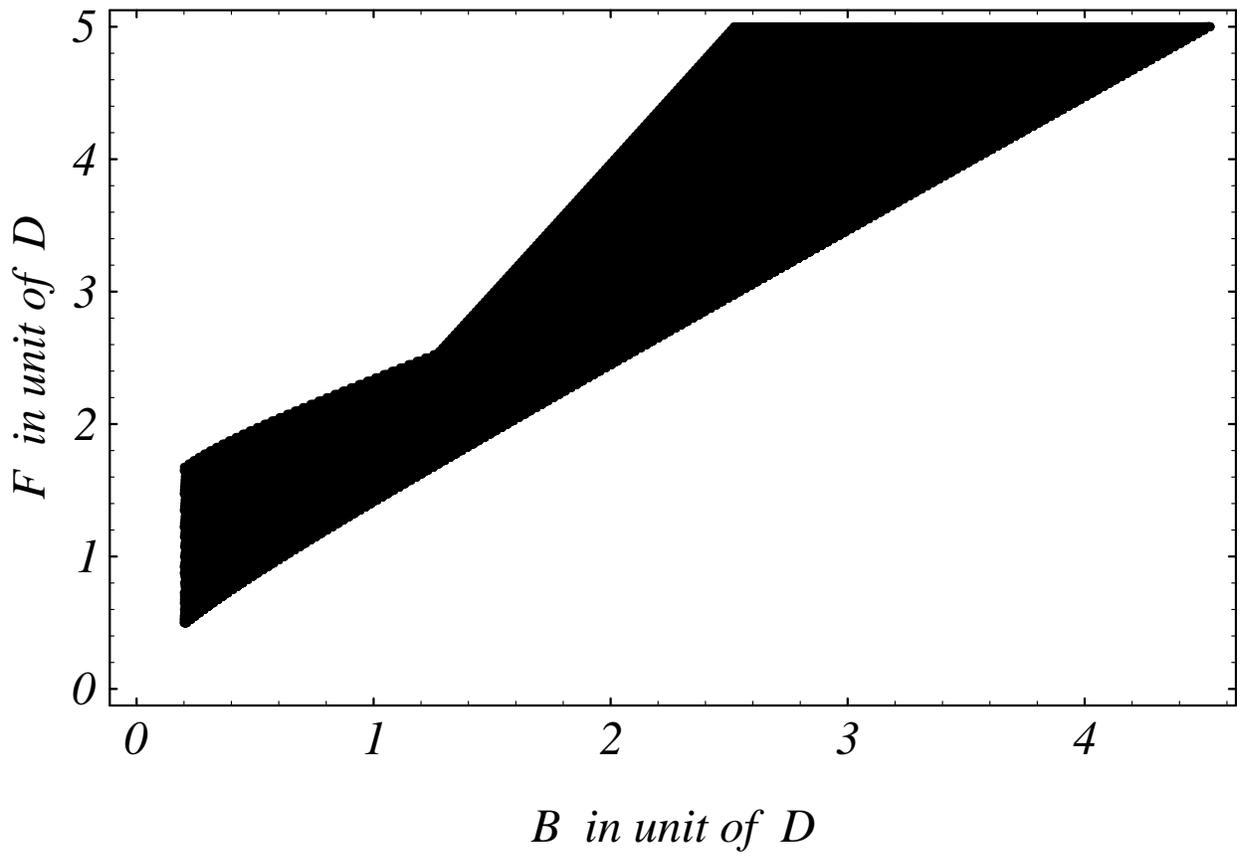
\end{document}